\documentclass[aps,prb,twocolumn,superscriptaddress,showpacs]{revtex4-1}
\usepackage[colorlinks=true,linkcolor=blue,filecolor=blue,menucolor=yellow,urlcolor=blue,citecolor=blue,anchorcolor=blue]{hyperref}
\usepackage{graphicx}% Include figure files
\begin{document}

% Use the \preprint command to place your local institutional report
% number in the upper righthand corner of the title page in preprint mode.
% Multiple \preprint commands are allowed.
% Use the 'preprintnumbers' class option to override journal defaults
% to display numbers if necessary
%\preprint{}

%Title of paper
\title{Graphene nanoribbons on vicinal SiC surfaces by molecular beam epitaxy}

% repeat the \author .. \affiliation  etc. as needed
% \email, \thanks, \homepage, \altaffiliation all apply to the current
% author. Explanatory text should go in the []'s, actual e-mail
% address or url should go in the {}'s for \email and \homepage.
% Please use the appropriate macro foreach each type of information

% \affiliation command applies to all authors since the last
% \affiliation command. The \affiliation command should follow the
% other information
% \affiliation can be followed by \email, \homepage, \thanks as well.
\author{Takashi Kajiwara}
%\homepage[]{Your web page}
%\thanks{}
%\altaffiliation{}
\author{Yuzuru Nakamori}
%\homepage[]{Your web page}
%\thanks{}
%\altaffiliation{}
\author{Anton Visikovskiy}
%\homepage[]{Your web page}
%\thanks{}
%\altaffiliation{}
\affiliation{Department of Applied Quantum Physics and Nuclear Engineering, Kyushu University, Fukuoka 819-0395, Japan}
\author{Takushi Iimori}
%\homepage[]{Your web page}
%\thanks{}
%\altaffiliation{}
\author{Fumio Komori}
%\homepage[]{Your web page}
%\thanks{}
%\altaffiliation{}
\affiliation{Institute of Solid State Physics, The University of Tokyo, Kashiwa 277-8581, Japan}
\author{Kan Nakatsuji}
%\homepage[]{Your web page}
%\thanks{}
%\altaffiliation{}
\affiliation{Department of Materials Science and Engineering, Tokyo Institute of Technology, Yokohama 226-8502, Japan}
\author{Kazuhiko Mase}
%\homepage[]{Your web page}
%\thanks{}
%\altaffiliation{}
\affiliation{Institute of Materials Structure Science, High Energy Accelerator Research Organization (KEK), Tsukuba 305-0801, Japan}
\author{Satoru Tanaka}
\email[stanaka@nucl.kyushu-u.ac.jp]{}
%\homepage[]{Your web page}
%\thanks{}
%\altaffiliation{}
\affiliation{Department of Applied Quantum Physics and Nuclear Engineering, Kyushu University, Fukuoka 819-0395, Japan}

%Collaboration name if desired (requires use of superscriptaddress
%option in \documentclass). \noaffiliation is required (may also be
%used with the \author command).
%\collaboration can be followed by \email, \homepage, \thanks as well.
%\collaboration{}
%\noaffiliation

\date{\today}

\begin{abstract}
We present a new method of producing a densely ordered array of epitaxial graphene nanoribbons (GNRs) using vicinal SiC surfaces as a template, which consist of ordered pairs of (0001) terraces and nanofacets. Controlled selective growth of graphene on approximately 10 nm wide of (0001) terraces with 10 nm spatial intervals allows GNR formation. By selecting the vicinal direction of SiC substrate, [1$\bar{1}$00], well-ordered GNRs with predominantly armchair edges are obtained. These structures, the high density GNRs, enable us to observe the electronic structure at K-points by angle-resolved photoemission spectroscopy, showing clear band-gap opening of at least 0.14 eV. 
\end{abstract}

% insert suggested PACS numbers in braces on next line
% insert suggested keywords - APS authors don't need to do this
%\keywords{}

%\maketitle must follow title, authors, abstract, \pacs, and \keywords
\maketitle

% body of paper here - Use proper section commands
% References should be done using the \cite, \ref, and \label commands
%\section{}
Graphene nanoribbons (GNRs) are attracting increasing attention in nanoelectronic applications and solid state physics, where the band-gap opening or modification of the electronic structure at K-points is a central interest \cite{Nakada1996}. The electronic structure at K-points in GNRs has theoretically\cite{Nakada1996, Barone2006, Yang2007} and experimentally\cite{Chen2007, Han2007, Wang2008, Kosynkin2009, Jiao2009, Li2008, Cai2010, Sprinkle2010} been demonstrated to depend on the type of edge geometry: armchair or zigzag. Semiconducting characteristics are expected in the case of armchair edges owing to the band-gap opening at K-points\cite{Barone2006}. As the width of GNRs is reduced, the gap is increased by both electron confinement and edge effects. However, realization of GNRs with atomically well-defined edges and providing experimental evidences of the gap opening at K-points remain challenging. Here we demonstrate a new approach for producing a densely ordered array of aligned GNRs, which is advantageous for investigating physical properties macroscopically, on unique SiC surfaces as templates via molecular beam epitaxy (MBE) and show band-gap openings at K-points visualized by angle-resolved photoemission spectroscopy (ARPES).
  Fabrication of GNRs by a conventional lithographic technique was first reported by Chen\cite{Chen2007} and Han\cite{Han2007}. GNRs as narrow as 15$ \sim $20 nm were obtained and the band-gap openings were shown; the band-gap value was dependent on the ribbon width. Since then, new and more advanced approaches have been proposed for GNR fabrication. These include unzipping of carbon nanotubes (CNTs)\cite{Kosynkin2009}, chemical\cite{Jiao2009}, sonochemical\cite{Li2008}, bottom-up molecular precursor methods\cite{Cai2010} and selective thermal decomposition at facets on SiC substrates\cite{Sprinkle2010}. Each method is unique and all have advantages and disadvantages. The unzipping of CNTs and chemical methods are appropriate for mass production but in their present state, require highly complicated treatment and lack sufficient quality control. The primary advantage of the molecular precursor method is that it enables very precise control of the width and edge structure; however, so far, it only works on metal surfaces. The thermal decomposition of SiC substrates, utilized in reference \onlinecite{Sprinkle2010}, is now recognized as a powerful method to obtain high-quality graphene with few structural defects\cite{Emtsev2009} owing to the epitaxial characteristic of the initial carbon layer with (6$\sqrt{3} $ $ \times $6$\sqrt{3} $)R30$ ^{\circ} $(6R3) registry on SiC(1$ \times $1)\cite{Emtsev2008, Kim2008}. GNRs, approximately 40 nm wide, were achieved by the thermal decomposition method with selective growth on facets created by reactive ion-etching on a SiC substrate. However, in this method, C atoms are supplied internally from SiC surface steps, and accordingly, unstable growth\cite{Ming2011, Ohta2010, Tanaka2010} and morphological fluctuation\cite{Sprinkle2010} are induced. Although all these methods have succeeded in forming GNRs of different quality, the physical characteristics of the resulting GNRs, in particular evident and direct observations of band-gap opening, have not been demonstrated.

\begin{figure}[b]
\includegraphics[scale = 0.54]{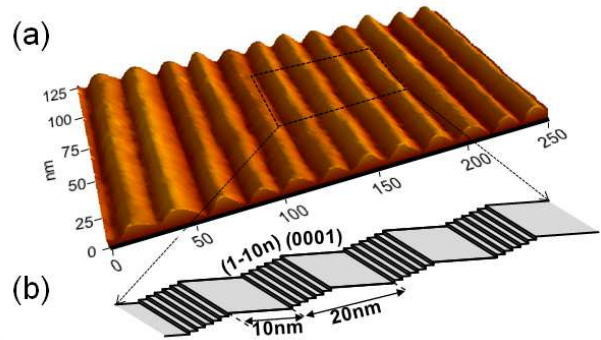}%
\caption{\label{fig:SiC}An AFM image and a structural model of SiC nanosurface.
(a) An AFM 3D view of the vicinal SiC surface after hydrogen etching, showing periodic array of terraces and facet structures. (b) A schematic drawing of the SiC surface in (a). Each pair of array consists of the (0001) terrace and (1$\bar{1}$0n) facet and is ordered with the periodic distance of $ \sim $20 nm. The width of each (0001) terrace is $ \sim $10 nm.
}
\end{figure}

   Here we report a new method for the fabrication of GNRs on a SiC substrate by molecular beam epitaxy (MBE). C atoms are supplied externally, which effectively addresses the problems described in reference  \onlinecite{Sprinkle2010}. Growth selectivity is achieved through the preferential nucleation of GNRs on terraces of vicinal SiC surfaces. The off-axis SiC substrate, often used in the fabrication of SiC electronic devices, is especially noted here because it shows a unique periodic surface structure. We have studied off-axis (vicinal) SiC surfaces that are intentionally off-cut from a (0001) plane and found self-ordered periodic structures consisting of pairs of a (0001) basal plane terrace and a (1$\bar{1}$0n) nanofacet (n = 35$ \sim $37) with a characteristic periodicity of $ \sim $20 nm, as shown in Fig.~\ref{fig:SiC}. The structural model, illustrated in Fig.~\ref{fig:SiC}(b), was derived from high-resolution transmission electron microscopy images.  The surface structure is formed by phase separation, quantized step-bunching and ordering\cite{Nakagawa2003}. Such a unique periodic SiC surface, hereafter SiC nanosurface, should be a good template for growing a massive array of GNRs. 

  In the present study, vicinal 6H-SiC substrates (Si-face, 4$ ^{\circ} $ off toward [1$\bar{1}$00]) first underwent H$ _{2} $ gas etching to produce a SiC nanosurface. The resultant SiC samples were then loaded into an ultra-high vacuum chamber, heated by direct current and cleaned under Si-flux at 1050$ ^{\circ} $C. In situ reflection high-energy electron diffraction (RHEED) indicated that the initial broad ($\sqrt{3} $$ \times $$\sqrt{3} $)R30$ ^{\circ} $(R3) structure of native oxide was changed to clear (3$ \times $3) of clean SiC(1$ \times $1) with Si adlayers\cite{Starke1998}, which are free from carbon and oxide contaminations. After terminating the Si-beam at the same temperature, Si adatoms R3 structure appeared owing to the evaporation of Si atoms in the Si adlayers\cite{Tautz2000}. The Si adatom R3 surface is important to achieve high-quality GNRs on SiC by MBE, because previous highly Si-rich (3$ \times $3) structure induces undesired SiC growth owing to residual Si atoms along with C atoms that are externally supplied, resulting in a collapse of the periodic morphology. After that, C atoms were supplied at a constant deposition rate by heating resistive carbon plates at 2200$ ^{\circ} $C. The RHEED pattern was monitored in situ at 15 keV and used to terminate the growth. Three samples were prepared by changing the growth time: 5, 20 and 33 min. Each sample was evaluated by atomic force microscopy (AFM), RHEED, low-energy electron diffraction (LEED) and Raman spectroscopy. Finally, the samples were exposed to hydrogen at 600$ ^{\circ} $C for 1 h to transform the 6R3 structure into quasi-free-standing graphene by hydrogen intercalation\cite{Riedl2009}. In addition, scanning tunneling microscopy (STM), polarized Raman spectroscopy and ARPES were performed. 

\begin{figure}
\includegraphics[scale = 0.50]{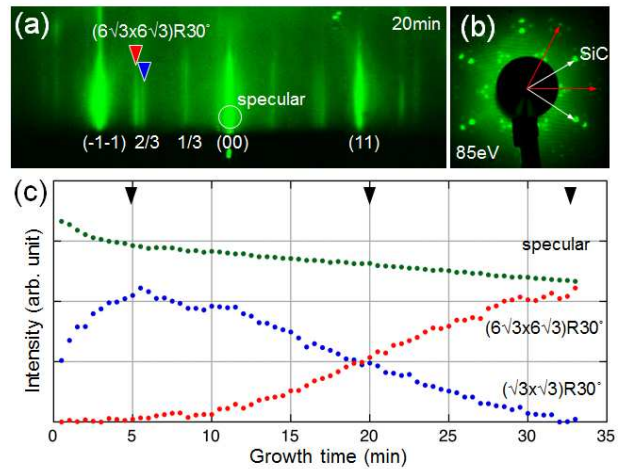}%
\caption{\label{fig:RHEED}In situ RHEED analysis during MBE growth and a LEED image after growth. (a) A RHEED image after 20 min growth, where both R3 and 6R3 structures are visible. (b) A LEED image after 33 min growth, indicating clear 6R3 satellite spots. (c) RHEED intensity profiles as a function of growth time. The intensities of a specular spot, an R3 and a 6R3 streak, as indicated in (a), are monitored. Note that the initial R3 intensity is increased after 5 min growth and the R3 and 6R3 intensities show a complementary relationship.
}
\end{figure}

  Figure~\ref{fig:RHEED}(a) shows the RHEED image after 20 min growth. The electron beam was irradiated parallel to the vicinal direction, i.e. perpendicular to step edges [1$\bar{1}$00]. Three sets of diffraction patterns are visible in this image: $ \times $1 SiC surface, $ \times $3 due to R3, and satellites due to 6R3. Figure 2(b) shows the evolution of RHEED intensity with growth time at selected positions; namely, a specular spot, R3 and 6R3 streaks, as indicated in Fig.~\ref{fig:RHEED}(a). The diffraction intensity of the R3 structure (2/3 streak) increased initially and decreased after $ \sim $5 min. The initial increase in the intensity is probably related to the structural change to the C-rich R3 surface. The detailed mechanism of structural transformation at this stage is not yet clear. We can speculate that the C atoms may incorporate into the Si-rich phase\cite{Righi2005} while maintaining the same surface geometry of R3 but resulting in higher intensity of the R3 streaks. This structure is now under investigation by in situ LEED I-V analysis. As the R3 intensity started to decrease, 6R3 streaks appeared and increased in intensity. This complementary relationship between the intensity profile of R3 and 6R3 structures implies structural transformation from R3 to 6R3 by the incorporation of C atoms. After 33 min growth, the transformation was complete; no R3 signal was visible. At this stage, the 6R3 layer covered the (0001) terraces, as shown in the LEED image in Fig.~\ref{fig:RHEED}(b). Note that no additional (overlap) spots except for SiC and the 6R3 structure are seen in this image, indicating that growth takes place selectively on (0001) terraces. The Raman spectrum of the sample grown for 33 min typically indicates no G' (2D) signal [see Fig.~\ref{fig:Raman}(a)], which also supports the formation of the 6R3 structure. 

\begin{figure}[htbp]
\includegraphics[scale = 0.48]{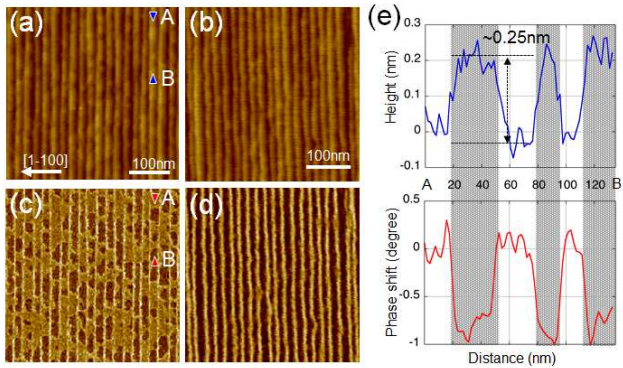}%
\caption{AFM analyses of GNRs. AFM height (upper) and phase images (lower) of the samples after 20 min growth (a and c) and 33 min growth (b and d) and the hydrogen intercalation. (e) represents the cross-section at the line A-B indicated in (a) and (c). $ \sim $0.25 nm higher regions shaded in (e) show darker contrast in the phase image, i.e. graphene on the hydrogen-terminated SiC surface.\label{fig:AFM}
}
\end{figure}

  Next, each sample was hydrogen intercalated to obtain quasi-free-standing graphene on the SiC surface (21). LEED (not shown) clearly indicated the absence of 6R3 satellites due to structural transformation and shows only superposition of clear (1$ \times $1) SiC and (1$ \times $1) graphene patterns. In the AFM height-mode, a slight contrast was visible on the (0001) terraces of the sample after 20 min growth, as shown in Fig.~\ref{fig:AFM}(a). No contrast was observed before the hydrogen treatment. This contrast results from the difference in height between graphene (previously 6R3) and the R3 regions. As the 6R3 area transformed to graphene, its height increased owing to hydrogen intercalation at the interface, whereas the R3 area was probably etched in the hydrogenation process. The height difference is $ \sim $0.25 nm, which is in good agreement with the cross-sectional height of graphene on the H-terminated SiC surface\cite{Markevich2012}. This is seen more clearly in Fig.~\ref{fig:AFM}(c), which shows the AFM phase image contrast for the corresponding area\cite{Hibino2010a} and is also evident from a comparison of cross-sectional height and phase profiles at the line A-B in Figs.~\ref{fig:AFM}(a) and (c). The sample grown for 33 min, i.e. covered with GNRs, shows a surface morphology that is very similar to that of the SiC nanosurface, as shown in Fig.~\ref{fig:AFM}(b), but areas of dark contrast at terraces are evident in the phase image, as shown in Fig.~\ref{fig:AFM}(d), owing to the formation of graphene. 

\begin{figure}
\includegraphics[scale = 0.46]{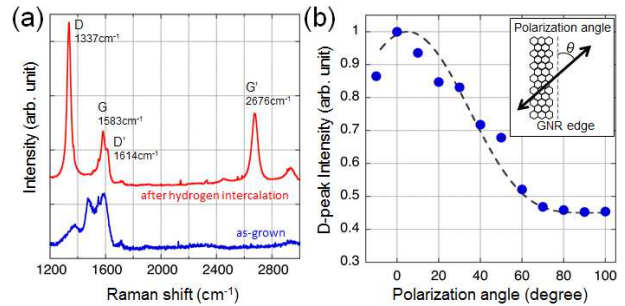}%
\caption{\label{fig:Raman}A Raman analyses of GNRs. (a) Raman spectra of the sample after 33 min growth before (blue) and after (red) hydrogen intercalation. (b) Polarization angle dependence of the D-band intensity. D-band intensity at each polarization angle $\theta $, defined in the insert, is fitted as a function of cos$ ^{4} $$\theta $ shown by a dotted curve.
}
\end{figure}

  The Raman spectra of the GNR sample (33 min growth) before and after hydrogen intercalation are shown in Fig.~\ref{fig:Raman}(a). The spectrum before hydrogen intercalation (as-grown) indicates typical features of the 6R3 structure, including lack of D- and G'-bands because of the absence of Dirac cones. In contrast, the GNR sample (after hydrogen intercalation) reveals evident D-, D'-, G-, and G'-bands. The D- and G'-bands show a fairly narrow peak fitted with a single Lorentzian function. The full-widths at half-maximum of the D and G' peaks are 31 and 49 cm$ ^{-1} $, respectively. The G-band peak at 1583 cm$ ^{-1} $ is due to LO phonons at armchair edges, which are softened owing to the Kohn anomaly effect\cite{Sasaki2009}. The deconvolution analysis of the G-band indicates that there is a small additional peak at a wavelength ~15 cm$ ^{-1} $ higher, which is probably due to either bulk LO phonons or TO phonons at armchair edges in the GNRs. The relatively strong D-band intensity due to the presence of high-density armchair edges and possible point defects should be noted. The edge character can be confirmed by the polarization angle dependence of the incident laser on the D-band intensity\cite{Cancado2004}. The integrated intensity of the D-band at polarization angle $\theta $ is plotted in Fig.~\ref{fig:Raman}(b). These are fitted as a function of cos$ ^{4} $$\theta $ , suggesting predominantly armchair-type edges. This is also supported by the STM observation; line nodes\cite{Yang2010} in the vicinity of the edges due to the interference effects of wave functions were recognized. 

\begin{figure}[b]
\includegraphics[scale = 0.50]{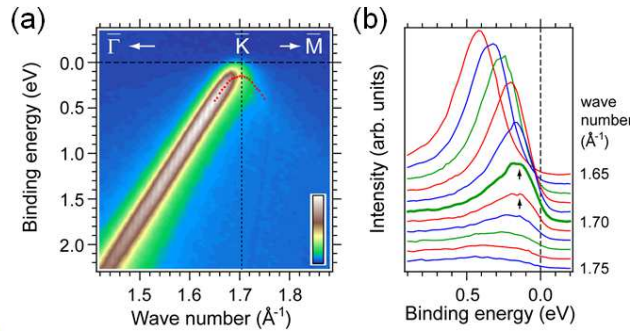}%
\caption{\label{fig:ARPES} Intensity map and EDCs of the ARPES spectra around the K-point. Intensity map (a) and EDCs (b) of the ARPES spectra around the K-point of the samples after 33 min growth. The spectra are taken along the $\Gamma $-K-M line. The red dots in (a) indicate the positions of the EDC peaks. The intensity map clearly indicates a linear dispersion of the valence band of the single-layer graphene. The EDCs show the folding of the valence band at the K-point below E$ _{F} $.
}
\end{figure}

  The band structures of the GNR samples were studied using ARPES. In Fig.~\ref{fig:ARPES}(a), the ARPES result along the $\Gamma $-K-M lines of the graphene surface Brillouin zone is shown as a photoemission intensity image. A linear valence band dispersion verifies the presence of a single layer graphene. Note that a conduction band is invisible and the band is folded at the K-point area, as shown in Fig.~\ref{fig:ARPES}(b). The red dots in Fig.~\ref{fig:ARPES}(a) indicate the peaks of the energy distribution curves (EDC) shown in Fig.~\ref{fig:ARPES}(b) and represent band dispersion around the K-point. No states are detected between the Fermi energy (E$ _{F} $) and valence band maximum, and a band-gap is opened at the K-point. The minimum band-gap of our GNRs can be 0.14 eV, assuming the conduction band minimum at E$ _{F} $. However, based on the results obtained in the hydrogen intercalation of the 6R3 sheet, the band-gap value is expected to be more than twice of 0.14 eV. A quasi-free-standing monolayer graphene shows a slight p-type doping nature\cite{Riedl2009, Ristein2012}, i.e. the Dirac point should be above E$ _{F} $.
  
  In summary, by templating vicinal SiC surfaces consisting of ordered pairs of (0001) terraces and nanofacets, GNRs are formed by MBE. The carbon atoms supplied to such a surface selectively organize a graphene network on (0001) terraces and form an epitaxial 6R3 layer. Hydrogen intercalation results in transformation to quasi-free-standing graphene and thus GNRs. The edge characteristic of the GNRs grown under these experimental conditions indicates armchair-type edges. The massive arrays of GNRs, $ \sim $10 nm in width, indicate apparent band-gap openings of at least 0.14 eV at the K-point observed by ARPES. A larger band-gap energy should be expected in GNRs $ \sim $5 nm in width. Such GNRs can be grown on vicinal 4H-SiC surfaces whose terrace width is $ \sim $5 nm\cite{Fujii2007}. These structures are also interesting in terms of catalytic phenomena, which require high density of edges to detect.
\begin{acknowledgments}
The authors would like to acknowledge Y. Hagihara for his technical contribution. This work was partly supported by a Grant in Aid for Scientific Research (KAKENHI, Grant No. 23246014) from the Ministry of Education, Culture, Sports, Science and Technology of Japan. ARPES measurements have been performed under the approval of the Photon Factory Program Advisory Committee (Proposal No. 2011G677).
\end{acknowledgments}

% Create the reference section using BibTeX:
\bibliography{bibliography}

\end{document}